\documentclass[aip,showpacs,preprint]{revtex4}
\usepackage{graphics}
\begin{document}
\title{Magnetocaloric properties of nanocrystalline 
La$_{0.125}$Ca$_{0.875}$MnO$_{3}$}


\author{Anis Biswas}
\email{anis.biswas@saha.ac.in}
\author{Tapas Samanta}
\email{tapas.samanta@saha.ac.in}
\author{S. Banerjee}
\author{I. Das}
\email{indranil.das@saha.ac.in}
\affiliation{Saha Institute of Nuclear Physics,1/AF,Bidhannagar,
Kolkata 700 064, India}

\begin{abstract}
Some recent experimental studies show the invisibility of antiferromagnetic 
transition in the cases of manganites when their particle size is reduced to 
nanometer scale. In complete contrast to these cases, we have observed the 
signature of antiferromagnetic transition in the magnetocaloric properties of 
nanocrystalline La$_{0.125}$Ca$_{0.875}$MnO$_{3}$ of average particle
 size $70$ and $60$ nm similar to its polycrystalline bulk form.
The system exhibit inverse magnetocaloric effect in its polycrystalline
 and nanocrystalline form. An extra ferromagnetic phase is stabilized 
at low temperature for the sample with particle size $\sim 60$ nm.    
\end{abstract}
\maketitle
 The physics of Perovskite Manganites has been a intriguing subject
 of intense research from both fundamental and application points of view
~\cite{tokura}.
 A recent trend of research in the field of manganites is to study
 the effect of the reduction of particle size on different properties
 of the materials~\cite{mahesh,ab1,zheng,ab2,rao1,rao2,sarkar1,dong1,dong2,ab3,
dey,ab4,ab5,luo,dou,ab6,ab7,ab8,wang}. 
It has already been reported that the particle
 size can have significant influence on the different phenomena exhibited
 by manganites~\cite{mahesh,ab1,zheng,ab2,rao1,rao2,sarkar1,dong1,dong2,ab3, dey,ab4,ab5,luo,dou,ab6,ab7,ab8,wang}. 
As a result, the transport, magnetic, magnetotransport and magnetocaloric
 properties of those materials are observed to be markedly modified when the
 particle size is reduced to the nanometer scale.
The main objective of the present study is to investigate the effect of
 the lowering of particle size on the magnetocaloric properties of 
La$_{0.125}$Ca$_{0.875}$MnO$_{3}$. In fact, the study of 
magnetocaloric properties of materials stimulates considerable research
 interest owing to its possible application in the magnetic refrigeration
~\cite{pecharsky1,phan,pecharsky2,ts1,ts2,ts3,ts4,ts5,nature}.

It has been reported that polycrystalline bulk 
La$_{0.125}$Ca$_{0.875}$MnO$_{3}$ shows paramagnetic
 to antiferromagnetic transition at $\sim$ $125$ K with the 
lowering of temperature~\cite{tokura}. There is influence of the
presence of inhomogeneous canted antiferromagnetic state on the 
antiferromagnetic transition for this compound~\cite{tokura}.
We have observed inverse magnetocaloric effect (IMCE) i.e, minimum in
 -$\Delta${S}(T) in the vicinity of 
antiferromagnetic (AFM) transition temperature for the polycrystalline bulk
 form of the sample. Such a minimum in -$\Delta${S}(T) is visible for its 
nanocrystalline forms around the same temperature as bulk also, which implies
 the occurring of antiferromagnetic transition for the nanoparticles of
 this manganite system. 
In addition to IMCE, a maximum in -$\Delta${S}(T) arises  at low temperate 
in the cases of
 nanocrystalline sample with the lowest particle size indicating stabilization 
of ferromagnetic state.
 The observation of clear signature of AFM transition for 
La$_{0.125}$Ca$_{0.875}$MnO$_{3}$ nanoparticles directly contradicts the
 previous experimental results of invisibility of AFM transition
 for many manganites due to the reduction of particle size
~\cite{ab1,rao1,ab4,rao2,sarkar1,dong1,dong2,luo}.          

The polycrystalline and nanocrystalline La$_{0.125}$Ca$_{0.875}$MnO$_{3}$ were 
prepared by solgel method. The details of the solgel method has been described in our previous article~\cite{ab1}. At the end of the solgel process, the
 decomposed gel was annealed at $1400^{o}$C for $36$ hours to prepare
 the polycrystalline sample. However, for preparing the nanoparticles of two 
different average particle size, the
 decomposed gel was subjected to heat treatment at $1200$ and $1000^{0}$ C for
 6 hours.  The x-ray powder
 diffraction study has confirmed the formation of the samples with single
 crystallographic phase. The particle size of the nanocrystalline
 samples was determined by Scanning Electron Microscopy (SEM) study. The
 average particle size of the nanocrystalline sample prepared at
 $1200^{0}$ C is $\sim$ $70$ nm. On the other hand,
 the average particle size of the sample synthesized at $1000^{0}$ C is
 $\sim$ $60$ nm. The two nanocrystalline samples are designated as
 S70 (sample with particle size $\sim 70$ nm) and S60 (sample with particle 
size $\sim 60$ nm) respectively. The typical SEM micrographs of the two
 samples have been given in Fig. 1.      

A commercial SQUID magnetometer was utilized for magnetization measurements.
The isothermal magnetic field dependence of magnetization [M(H)] at different
 temperatures has been studied for the samples (Fig. 2).
From  the isothermal M(H) curves, the change of the magnetic entropy
(-$\Delta{S}$) was estimated for
various magnetic fields by using the Maxwell's relation~\cite{ts1},
\begin{equation}
\left(\frac{\partial{S}}{\partial{\mu_{0} H}}\right)_{T} = \left( \frac{\partial{M}}{\partial{T}}\right)_{\mu_{0} H}
\end{equation}
 The temperature dependence of -$\Delta{S}$ for $7$ T
 has been shown in Fig. 3. A minimum in -$\Delta{S}$(T) 
(inverse magnetocaloric effect) has been
 observed at $\sim 125$ K for the polycrystalline as well as the 
nanocrystalline samples. The materials exhibiting IMCE would be very useful
 for magnetic refrigeration as the refrigerant capacity can be enhanced
 by utilizing those materials in composite with the conventional
 magnetic refrigerants~\cite{nature}. However the manifestation of IMCE
 in the cases of manganites is rare.  
For the polycrystalline 
La$_{0.125}$Ca$_{0.875}$MnO$_{3}$, antiferromagnetic
 transition occurs at $\sim 125$ K~\cite{tokura}. In the cases of 
antiferromagnetic 
systems, when the applied magnetic field is below the field
 required for the quenching of antiferromagnetism, the IMCE can be observed
~\cite{tishin}. 
The IMCE for the present polycrystalline sample originates due to the
 antiferromagnetic transition. Similarly, the minimum in -$\Delta${S}(T) around
 the same temperature as the polycrystalline sample
 for S70 and S60 can be considered as the signature of the 
antiferromagnetic transition. There are number of studies on different
 nanocrystalline systems, which reveal that the antiferromagnetic transition
 in bulk system becomes invisible due to the reduction of particle size
 ~\cite{ab1,rao1,ab4,rao2,sarkar1,dong1,dong2,luo}.
The present result for the nanocrystalline La$_{0.125}$Ca$_{0.875}$MnO$_{3}$,
 contradicts those existing results.
The stabilization of antiferromagnetic state in the previous systems is mostly
 of C-E type and arises as because of the occurrence of charge order 
transition (COT). In their nanocrystalline form, COT does not occur resulting in
 the hindrance of antiferromagnetic transition
~\cite{ab1,rao1,ab4,rao2,sarkar1,dong1,dong2,luo,dou}. The phase diagram
 of La$_{1-x}$Ca$_{x}$MnO$_{3}$ reveals that 
La$_{0.125}$Ca$_{0.875}$MnO$_{3}$ situates at the phase boundary and charge
 ordering correlation is hardly present in this compound~\cite{tokura}. 
The antiferromagnetic state originated in this compound is mixture of
 C-type and G-type. From the present result, it seems that because of 
different type of antiferromagnetic 
 ordering for La$_{0.125}$Ca$_{0.875}$MnO$_{3}$ in comparison to the 
charge ordered antiferromagnetic
 state for previous cases, the antiferromagnetic transition can be
 visible in the nanocrystalline form of this material.      
An additional feature of -$\Delta$S(T) for S60 is a clear
 maximum in low temperature. Such a maximum is an indication of
 the onset of ferromagnetism for that nanocrystalline sample.
Very recent theoretical study predicts that the ferromagnetic correlation
 can be originated due to the surface effect in the cases of fine
 particles of manganites~\cite{dong2}. The signature of ferromagnetism at low 
temperature
 in the
-$\Delta$S(T) observed for
S60  can be attributed to the surface effect in accordance to the theoretical
 prediction~\cite{dong2}. 
For the nanocrystalline samples, the surface effect becomes more dominant
 as the particle size of the sample reduces. It appears
 that the surface effect for S$70$ is not so enough to create ferromagnetic 
state. However in the case of 
S$60$ with lower particle size, surface effect can be more
 dominant giving rise to ferromagnetism at low temperature.
The M-H curves  at $45$ K and $95$ K for the two samples are also different.
There is a cross over between M-H curves at $45$ K and $95$ K in the 
low magnetic field region for S$70$ like its bulk form. This cross over
 between two M-H curves is absent for S$60$ (Fig. 2).

To summarize, we have observed IMCE for polycrystalline as well as 
nanocrystalline La$_{0.125}$Ca$_{0.875}$MnO$_{3}$ due to the antiferromagnetic
 transition. The evidence of antiferromagnetic transition in the
 cases of nanocrystalline manganites is in direct contradiction of
 existing experimental results for different manganite systems ~\cite{ab1,rao1,ab4,rao2,sarkar1,dong1,dong2,luo,dou}. Reduction of particle size has also
 pronounced effect on the magnetocaloric properties of this material as
 an additional maximum is visible in -$\Delta$S(T) for the sample with average
 particle size $60$ nm. The observation of IMCE is very rare
 especially in the cases of rare-earth based manganites.

\newpage
\noindent
FIG.1: Scanning electron micrograph for nanocrystalline 
La$_{0.125}$Ca$_{0.875}$MnO$_{3}$ of  average particle size (a) $70$ nm and
 (b) $60$ nm. The samples of average particle size $\sim 70$ nm
 and $\sim 60$ nm are designated as S$70$ and S$60$ respectively. \\

FIG. 2: M versus H curves at some temperatures for (a) 
polycrystalline bulk (b) S$70$ and (c) S$60$.  M(H) curves at other 
temperatures are
 not shown because of sake of clarity.\\

FIG. 3: The temperature dependence of -$\Delta{S}$ for $7$ T in
 the cases of (a) polycrystalline bulk, (b) S$70$ and (c) S$60$.




\newpage
\noindent
\begin{figure}
\resizebox{8.0cm}{!}
{\includegraphics{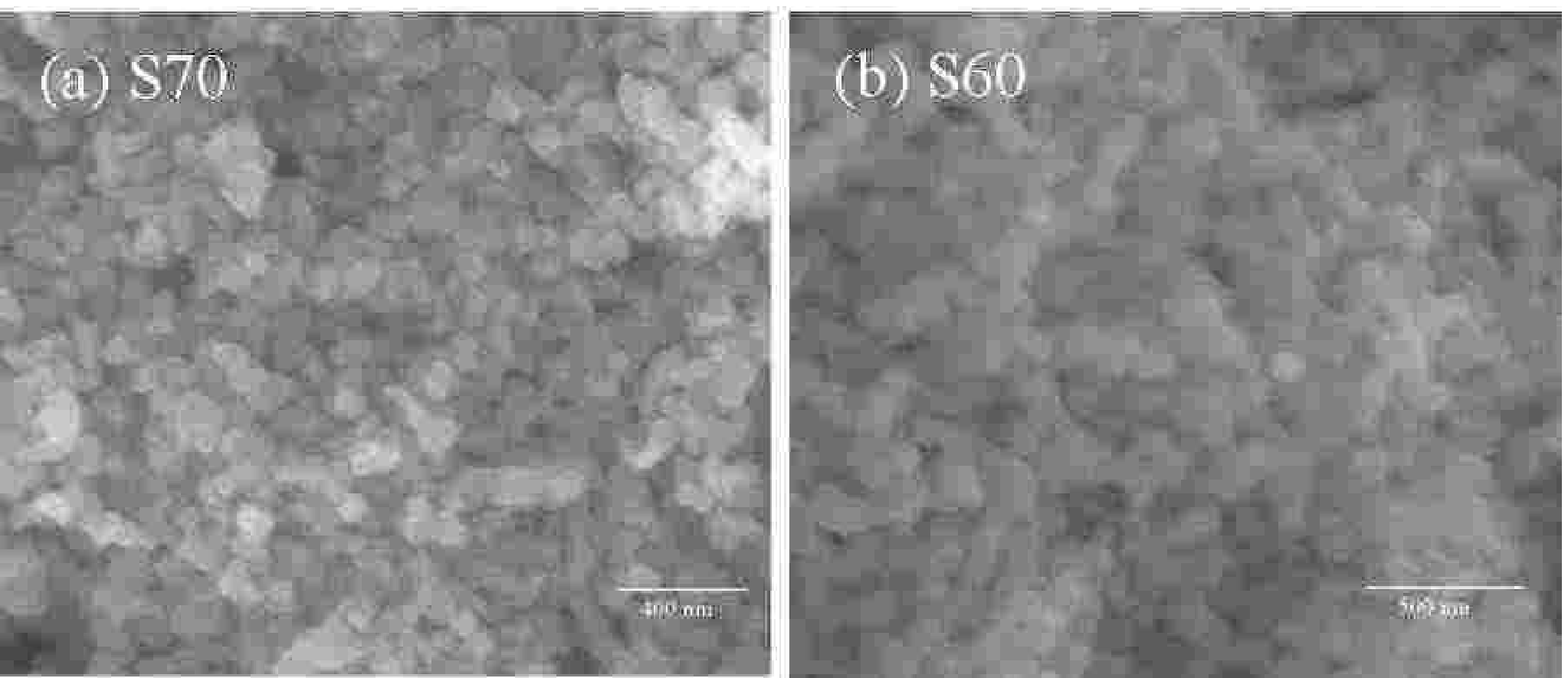}}
\caption{}
\end{figure}

\newpage
\begin{figure}
\resizebox{6.0cm}{!}
{\includegraphics{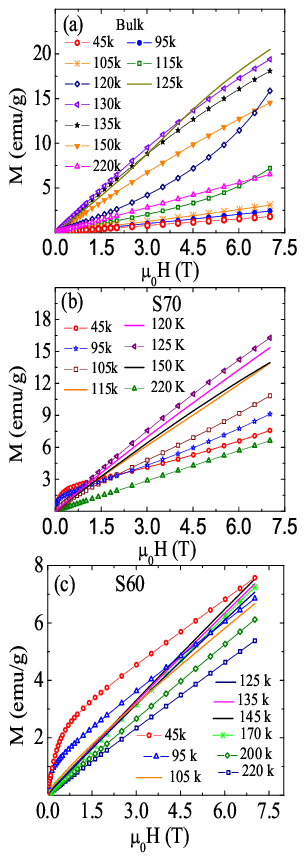}}
\caption{}
\end{figure}

\newpage
\begin{figure}
\resizebox{6.0cm}{6cm}
{\includegraphics{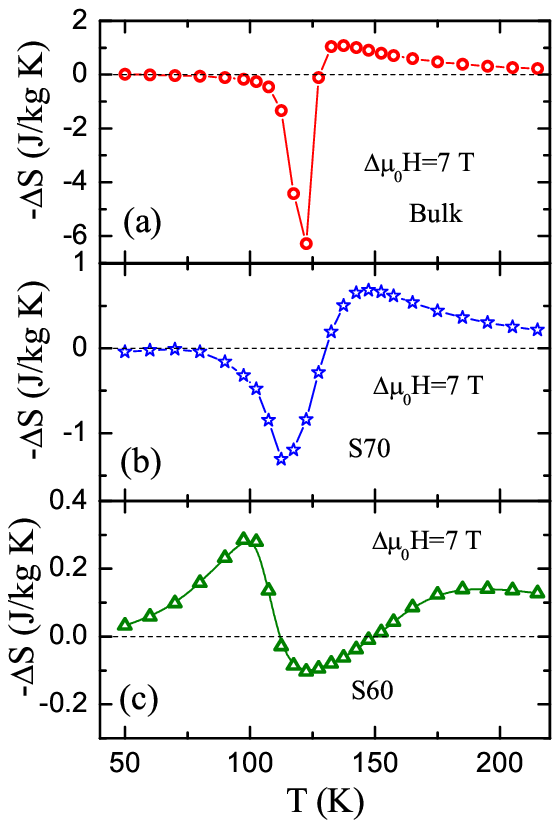}}
\caption{}
\end{figure}

\end{document}